\documentclass[12pt]{article}
\title{
\vspace{-3mm}
\rightline{\small IFUP-TH 2002/27}
\vspace{8mm}
\bf String representation of the SU(N)-inspired dual Abelian-Higgs--type theory with the $\Theta$-term}
\author{Dmitri Antonov
\thanks{E-mail: {\tt antonov@df.unipi.it}}
\\
{\it INFN-Sezione di Pisa, Universit\'a degli studi di Pisa,}\\
{\it Dipartimento di Fisica, Via Buonarroti, 2 - Ed. B - 56127 Pisa, Italy}\\
and\\
{\it Institute of Theoretical and Experimental Physics,}\\
{\it B. Cheremushkinskaya 25, RU-117 218 Moscow, Russia}}
\date{}
\begin{document}
\maketitle
\vspace{1mm}
\centerline{\bf {Abstract}}
\vspace{3mm}
\noindent
String representation of the $[U(1)]^{N-1}$ gauge-invariant
dual Abelian-Higgs--type theory, which is relevant to the $SU(N)$-QCD
with the $\Theta$-term and provides confinement of quarks, is derived.
The $N$-dependence of the Higgs vacuum expectation value is found, at which
the tension of the string joining quarks becomes $N$-independent, similarly to the real QCD. Contrary to that,
the inverse coupling constant of the rigidity term of this string always behaves approximately as $1/N$.
A long-range Aharonov-Bohm--type interaction of a dyon (i.e., a quark which acquired a magnetic charge
due to the $\Theta$-term) with a closed electric string becomes nontrivial at
$\Theta\ne N\pi\times{\,}{\rm integer}$. On the contrary, at these critical values of $\Theta$,
the scattering of dyons over strings is absent.

\vspace{3mm}

\vspace{3mm}

\newpage

\section{Introduction. The model.}

During the last years,
the method of Abelian projections~\cite{th} has been extensively used
both analytically and numerically to describe confinement
in QCD by the monopole mechanism (for recent reviews see~\cite{digiacomo} and refs. therein).
In particular, several attempts have been done to address the case of arbitrary number of colors~\cite{suN, suNN}.
On the way of using the method of Abelian projections, it is reasonable to base the respective 3D {\it continuum} models on the
assumption that monopoles form a dilute plasma (see e.g. ref.~\cite{dw} for the $SU(2)$-case).
This is because such a monopole configuration is an approximate stationary point of the
action of the $SU(N)$ 3D Georgi-Glashow model, and the confining mechanism of the latter is supposed to be
similar to that of Abelian-projected theories~\cite{th}.
In the present letter, we shall work in 4D and explore another
$SU(N)$-inspired theory describing Abelian-projected monopoles, which provides confinement of quarks.
It is based on the alternative assumption~\cite{tHM} that monopoles form magnetic
Higgs condensate, rather than the plasma. This assumption looks more appropriate in 4D, where Abelian-projected monopoles are known to be
proliferating~\cite{pb}, and therefore cannot be treated in the approximation of a dilute plasma.
The model we are going to deal with is a straightforward generalization of the respective $SU(3)$-one~\cite{maedan},
whose string representation
has been explored in refs.~\cite{su3, theta} (see also~\cite{moresu3} where the collective effects of vortex loops in this model
have been studied).
Similarly to ref.~\cite{theta}, we shall consider the general case of a
theory extended by the $\Theta$-term, owing to which quarks acquire a nonvanishing magnetic charge (i.e., become dyons)
and scatter over the dual electric Abrikosov-Nielsen-Olesen strings~\cite{ano}. Note that the simplest model
of this type, corresponding to the Abelian-projected $SU(2)$-QCD with the $\Theta$-term, has for the first time
been considered in ref.~\cite{emil}. As one of the results of the present letter, we shall get the
critical values of $\Theta$ in the $SU(N)$-case, at which the long-range topological interaction of dual strings
with dyons disappears. These values in particular reproduce the respective $SU(2)$- and $SU(3)$-ones, obtained in the above-mentioned
papers.

The partition function of the
effective $[U(1)]^{N-1}$ gauge-invariant Abelian-projected theory
we are going to explore~\footnote{
Throughout the present letter,
all the investigations will be performed in the Euclidean space-time.}
reads

$$
{\cal Z}_\alpha=\int\left(\prod\limits_{i}^{} \left|\Phi_i\right| {\cal D}\left|\Phi_i\right|
{\cal D}\theta_i\right) {\cal D}{\bf B}_\mu
\delta\left(\sum\limits_{i}^{}
\theta_i\right)\exp\Biggl\{-\int d^4x\Biggl[\frac14\left({\bf F}_{\mu\nu}+{\bf F}_{\mu\nu}^{(\alpha)}\right)^2+$$

\begin{equation}
\label{et6}
+\sum\limits_{i}^{}\left[\left|\left(\partial_\mu-
ig_m{\bf q}_i{\bf B}_\mu\right)\Phi_i\right|^2+
\lambda\left(|\Phi_i|^2-\eta^2\right)^2\right]-\frac{i\Theta g_m^2}{16\pi^2}
\left({\bf F}_{\mu\nu}+{\bf F}_{\mu\nu}^{(\alpha)}\right)
\left(\tilde{\bf F}_{\mu\nu}+\tilde{\bf F}_{\mu\nu}^{(\alpha)}\right)
\Biggr]\Biggr\}.
\end{equation}
Here, the index $i$ runs from 1 to the number of positive roots ${\bf q}_i$'s of the $SU(N)$-group, that is $N(N-1)/2$.
Next, $g_m$ is the magnetic
coupling constant related to the
electric one, $g$, by means of the topological
quantization condition $g_mg=4\pi n$. In what follows,
we shall for simplicity restrict ourselves to the monopoles
possessing the minimal charge only, i.e., set $n=1$, although the generalization to an
arbitrary $n$ is straightforward.
Note that the origin of root vectors in eq.~(\ref{et6}) is the fact that
monopole charges are distributed along them. Further,
$\Phi_i=\left|\Phi_i\right|{\rm e}^{i\theta_i}$ are the
dual Higgs fields, which describe the condensates of monopoles, and
${\bf F}_{\mu\nu}=\partial_\mu{\bf B}_\nu-\partial_\nu{\bf B}_\mu$ is the
field-strength tensor of the
$(N-1)$-component ``magnetic'' potential ${\bf B}_\mu$. The latter is dual
to the ``electric'' potential, whose components are diagonal gluons.
Since the $SU(N)$-group is special, the phases $\theta_i$'s of the
dual Higgs fields are related to each other by the constraint
$\sum\limits_{i}^{}\theta_i=0$, which is imposed by introducing
the corresponding $\delta$-function into the r.h.s. of eq.~(\ref{et6}).
Next, the index $\alpha$ runs from 1 to $N$ and denotes a certain quark color.
Finally, $\tilde {\cal O}_{\mu\nu}\equiv\frac12\varepsilon_{\mu\nu\lambda\rho}
{\cal O}_{\lambda\rho}$, and
${\bf F}_{\mu\nu}^{(\alpha)}$ is the field-strength tensor
of a test quark of the color $\alpha$, which moves along a certain
contour $C$. This tensor obeys the equation
$\partial_\mu\tilde {\bf F}_{\mu\nu}^{(\alpha)}=g{\bf m}_\alpha j_\nu$,
where $j_\mu(x)=\oint\limits_{C}^{}dx_\mu(\tau)\delta(x-x(\tau))$,
and ${\bf m}_\alpha$ is a weight vector of the group $SU(N)$.
One thus has ${\bf F}_{\mu\nu}^{(\alpha)}=g{\bf m}_\alpha\tilde{\cal F}_{\mu\nu}$,
where ${\cal F}_{\mu\nu}$ can be chosen e.g. in the form
${\cal F}_{\mu\nu}=-\Sigma_{\mu\nu}$.
Here, $\Sigma_{\mu\nu}(x)=\int\limits_{\Sigma}^{}d\sigma_{\mu\nu}
(x(\xi))\delta(x-x(\xi))$ is the vorticity tensor current associated
with the world sheet $\Sigma$ of
the open electric string, bounded by the contour $C$~\footnote{Another possible choice of ${\cal F}_{\mu\nu}$
is ${\cal F}_{\mu\nu}(x)=\partial_\nu^x\int d^4yD_0(x-y)j_\mu(y)-(\mu\leftrightarrow\nu)$, where $D_0(x)=1/(4\pi^2x^2)$
is the massless propagator. The obvious difference between these two choices is the dimensionality of the support
of ${\cal F}_{\mu\nu}$ -- either it is a 2D Dirac sheet $\Sigma$, or the whole 4D space-time.
It is known, however, that this ambiguity in the choice of the solution to the equation $\partial_\mu{\cal F}_{\mu\nu}=j_\nu$
does not affect physical results.}.
From now on, we shall omit the normalization constant in front of
all the functional integrals implying for every color $\alpha$ the normalization condition
${\cal Z}_\alpha\left[C=0\right]=1$.

Note that the $\Theta$-term can be rewritten as

\begin{equation}
\label{ch}
-\frac{i\Theta g_m^2}{16\pi^2}
\left({\bf F}_{\mu\nu}+{\bf F}_{\mu\nu}^{(\alpha)}\right)
\left(\tilde{\bf F}_{\mu\nu}+\tilde{\bf F}_{\mu\nu}^{(\alpha)}\right)=\frac{i\Theta g_m}{\pi}{\bf m}_\alpha
\int d^4x{\bf B}_\mu j_\mu,
\end{equation}
which means that by virtue of this term quarks start interacting with the magnetic gauge field ${\bf B}_\mu$~\cite{witten}.
This is only possible provided they
acquire some magnetic charge, i.e., become dyons. According to eq.~(\ref{ch}), this charge is indeed nonvanishing and
equals to $\Theta g_m/\pi$.

Expanding for a while $|\Phi_i|$ around the Higgs v.e.v. $\eta$, one gets the mass of the dual vector boson,
$m=g_m\eta\sqrt{N}$. In what follows, we shall work in the London limit of the model~(\ref{et6}),
which admits a construction of the string representation. This is the limit when $m$ is much smaller
than the mass of any of the Higgs fields, $m_H=\eta\sqrt{2\lambda}$. Since we would like
the model under study be consistent with QCD, we must have $g\sim\sqrt{\bar\lambda/N}$, where $\bar\lambda$
is the 't Hooft coupling constant, which remains finite in the large-$N$ limit. Therefore, in the London limit,
the Higgs coupling $\lambda$ should grow with $N$ faster than ${\cal O}\left(N^2\right)$, namely it should obey the
inequality $\lambda\gg 8\pi^2N^2/\bar\lambda$.

Integrating $|\Phi_i|$'s out, we arrive at the following expression for the partition function~(\ref{et6})
in the London limit:

$$
{\cal Z}_\alpha=\int\left(\prod\limits_{i}^{}
{\cal D}\theta_i^{\rm sing}{\cal D}\theta_i^{\rm reg}\right) {\cal D}{\bf B}_\mu{\cal D}k
\delta\left(\sum\limits_{i}^{}
\theta_i^{\rm sing}\right)\exp\Biggl\{-\int d^4x\Biggl[\frac14\left({\bf F}_{\mu\nu}+{\bf F}_{\mu\nu}^{(\alpha)}\right)^2+$$

\begin{equation}
\label{et7}
+\eta^2\sum\limits_{i}^{}\left(\partial_\mu\theta_i-
g_m{\bf q}_i{\bf B}_\mu\right)^2-ik\sum\limits_{i}^{}\theta_i^{\rm reg}-\frac{i\Theta g_m^2}{16\pi^2}
\left({\bf F}_{\mu\nu}+{\bf F}_{\mu\nu}^{(\alpha)}\right)
\left(\tilde{\bf F}_{\mu\nu}+\tilde{\bf F}_{\mu\nu}^{(\alpha)}\right)
\Biggr]\Biggr\}.
\end{equation}
Here, we have decomposed the total phases
of the dual Higgs
fields into multivalued and singlevalued (else oftenly called singular and
regular, respectively) parts, $\theta_i=
\theta_i^{\rm sing}+
\theta_i^{\rm reg}$, and imposed the constraint of vanishing of the
sum of regular parts by introducing the integration over the
Lagrange multiplier $k(x)$. The fields
$\theta_i^{\rm sing.}$'s
describing a certain configuration of closed dual strings
are related to the world sheets
$\Sigma_i$'s of these strings
by means of the equation

\begin{equation}
\label{suz3}
\varepsilon_{\mu\nu\lambda\rho}\partial_\lambda\partial_\rho
\theta_i^{\rm sing}(x)=2\pi\Sigma_{\mu\nu}^i(x)\equiv
2\pi\int\limits_{\Sigma_i}^{}d\sigma_{\mu\nu}\left(x^{(i)}(\xi)\right)
\delta\left(x-x^{(i)}(\xi)\right).
\end{equation}
This equation is the covariant formulation
of the 4D analogue of the Stokes' theorem for the gradient
of the field $\theta_i$, written in the local form. In eq.~(\ref{suz3}),
$x^{(i)}(\xi)\equiv x_\mu^{(i)}(\xi)$ is a vector, which parametrizes
the world sheet $\Sigma_i$ with $\xi=(\xi^1, \xi^2)$
standing for the 2D coordinate. As far as
the regular parts of the phases, $\theta_i^{\rm reg}$'s, are concerned,
those describe single-valued fluctuations around the string configuration described by $\theta_i^{\rm sing}$'s.
Note that owing to the one-to-one correspondence between
$\theta_i^{\rm sing}$'s and $\Sigma_i$'s, established by
eq.~(\ref{suz3}),
the integration over $\theta_i^{\rm sing}$'s is implied
in the sense of a certain prescription of the summation
over string world sheets. For the $SU(3)$-inspired model, one of the possible
concrete forms of such a prescription, corresponding to the summation
over the grand canonical ensemble of
virtual pairs of strings with opposite winding numbers, has been
considered in ref.~\cite{moresu3}.
It is also worth noting that by virtue of eq.~(\ref{suz3}) it is possible to demonstrate that
the integration measure ${\cal D}\theta_i$ becomes
factorized into the product ${\cal D}\theta_i^{\rm sing}{\cal D}\theta_i^{\rm reg}$.

\section{String representation.}

Let us now construct the string representation of the model~(\ref{et7}).
First, similarly to the $SU(3)$-case~\cite{su3, theta}, one can show that due to the equality $\sum\limits_{i}^{}{\bf q}_i=0$,
the integration over $k$ yields only an inessential constant factor, and we get

$$
\int\left(\prod\limits_{i}^{}
{\cal D}\theta_i^{\rm sing}
{\cal D}\theta_i^{\rm reg}\right) {\cal D}k\delta\left(\sum\limits_{i}^{}
\theta_i^{\rm sing}\right)
\exp\Biggl\{-\int d^4x\Biggl[\eta^2
\sum\limits_{i}^{}\left(\partial_\mu\theta_i-
g_m{\bf q}_i{\bf B}_\mu\right)^2-ik\sum\limits_{i}^{}\theta_i^{\rm reg}\Biggr]\Biggr\}=$$

$$
=\int\left(\prod\limits_{i}^{}{\cal D}x^{(i)}(\xi){\cal D}h_{\mu\nu}^i\right)
\delta\left(\sum\limits_{i}^{}\Sigma_{\mu\nu}^i\right)
\exp\Biggl\{-\int d^4x\Biggl[\frac{1}{24\eta^2}
\left(H_{\mu\nu\lambda}^i\right)^2-i\pi h_{\mu\nu}^i\Sigma_{\mu\nu}^i+
ig_m{\bf q}_i
{\bf B}_\mu\partial_\nu \tilde h_{\mu\nu}^i\Biggr]\Biggr\}.
$$
Here, the Kalb-Ramond field $h_{\mu\nu}^i$ is dual to $\theta_i^{\rm reg}$, and
$H_{\mu\nu\lambda}^i=\partial_\mu h_{\nu\lambda}^i+
\partial_\lambda h_{\mu\nu}^i+\partial_\nu h_{\lambda\mu}^i$ stands
for the strength tensor of this field.
We have also used the relation~(\ref{suz3})
and referred the Jacobians~\cite{polikarp} emerging in course of the change of variables $\theta_i^{\rm sing}\to x^{(i)}$
to the integration measures ${\cal D}x^{(i)}(\xi)$'s.

The action of the dual-gauge-field sector of the model can then be written as follows:

$$\int d^4x\Biggl[
\frac14{\bf F}_{\mu\nu}^2+\frac14\left({\bf F}_{\mu\nu}^{(\alpha)}\right)^2+
{\bf B}_\mu\partial_\nu\left(
ig_m{\bf q}_i\tilde h_{\mu\nu}^i-
g{\bf m}_\alpha\tilde\Sigma_{\mu\nu}-
\frac{i\Theta g_m^2}{4\pi^2}\tilde{\bf F}_{\mu\nu}^{(\alpha)}\right)\Biggr].$$
The ${\bf B}_\mu$-fields can then be integrated out as Lagrange multipliers by passing to the
new fields $B_\mu^i={\bf q}_i{\bf B}_\mu$, using the formula~\cite{group}~\footnote{See also the last paper in ref.~\cite{suNN}
for the discussion of this formula.} $\left(B_\mu^i\right)^2=\frac{N}{2}{\bf B}_\mu^2$,
and introducing the numbers $s_i^{(\alpha)}$'s
according to the definition ${\bf m}_\alpha={\bf q}_is_i^{(\alpha)}$. The resulting partition function reads
as follows:

$$
{\cal Z}_\alpha=\int\left(\prod\limits_{i}^{}{\cal D}x^{(i)}(\xi){\cal D}h_{\mu\nu}^i\right)
\delta\left(\sum\limits_{i}^{}\Sigma_{\mu\nu}^i\right)
\exp\Biggl\{-\int d^4x\Biggl[\frac{1}{24\eta^2}
\left(H_{\mu\nu\lambda}^i\right)^2-i\pi h_{\mu\nu}^i\Sigma_{\mu\nu}^i+$$

\begin{equation}
\label{zA}
+\frac{N}{8}\left(
g_mh_{\mu\nu}^i+igs_i^{(\alpha)}
\Sigma_{\mu\nu}-\frac{\Theta g_m}{\pi} s_i^{(\alpha)}
\tilde{\cal F}_{\mu\nu}\right)^2
+\frac14\left({\bf F}_{\mu\nu}^{(\alpha)}
\right)^2\Biggr]\Biggr\}.
\end{equation}

To proceed with the analysis of this expression, we obviously need to know possible values of $s_i^{(\alpha)}$'s,
as well as $\left(s_i^{(\alpha)}\right)^2$ for a fixed $\alpha$. First of all,
it is straightforward to see that
for a given $\alpha$, only $(N-1)$ numbers $s_i^{(\alpha)}$'s are different from zero.
This is simply because
only $(N-1)$ ${\bf q}_i$'s out of $N(N-1)/2$ positive roots are so that ${\bf m}_\alpha{\bf q}_i=1/2$,
while the others are orthogonal to ${\bf m}_\alpha$.
Next, by noting that every root vector can be represented as a difference of two weight vectors and by using the
normalization condition ${\bf m}_\alpha{\bf m}_\beta=\left(\delta_{\alpha\beta}-N^{-1}\right)/2$, these nonvanishing $s_i^{(\alpha)}$'s can be found
to be $\pm N^{-1}$ (with $\sum\limits_{i}^{}s_i^{(\alpha)}=0$),
so that $\left(s_i^{(\alpha)}\right)^2=(N-1)/N^2$. Owing to this result, the singular term
$\frac14\left({\bf F}_{\mu\nu}^{(\alpha)}\right)^2$ in eq.~(\ref{zA}) cancels out, and we get the following intermediate expression
for the partition function:

$$
{\cal Z}_\alpha=\exp\left[-\frac{N-1}{8N}\left(\frac{\Theta g_m}{\pi}\right)^2\int d^4x
{\cal F}_{\mu\nu}^2
-\frac{2i\Theta(N-1)}{N}\hat L(\Sigma, C)\right]\times$$

\begin{equation}
\label{result}
\times\int\left(\prod\limits_{i}^{}{\cal D}x^{(i)}(\xi){\cal D}h_{\mu\nu}^i\right)
\delta\left(\sum\limits_{i}^{}\Sigma_{\mu\nu}^i\right)\exp\Biggl\{-\int d^4x\Biggl[\frac{1}{24\eta^2}
\left(H_{\mu\nu\lambda}^i\right)^2+\frac{Ng_m^2}{8}\left(h_{\mu\nu}^i\right)^2
-i\pi h_{\mu\nu}^i\Sigma_{\mu\nu}^{i{\,}(\alpha)}\Biggr]\Biggr\}.
\end{equation}
Here, $\hat L(\Sigma, C)\equiv\int d^4xd^4y\tilde\Sigma_{\mu\nu}(x)j_\nu(y)\partial_\mu^xD_0(x-y)$
is the (formal expression for the) 4D Gauss' linking number
of the surface $\Sigma$ with its boundary $C$, which eventually
becomes cancelled from the
final expression for ${\cal Z}_\alpha$, and
$\Sigma_{\mu\nu}^{i{\,}(\alpha)}\equiv\Sigma_{\mu\nu}^i-
Ns_i^{(\alpha)}\Sigma_{\mu\nu}-
\frac{i\Theta Ng_m^2}{4\pi^2}s_i^{(\alpha)}\tilde{\cal F}_{\mu\nu}$, so that
$\partial_\mu\Sigma_{\mu\nu}^{i{\,}(\alpha)}=Ns_i^{(\alpha)}j_\nu$.

Further integration over the Kalb-Ramond fields is straightforward and yields

$$\int\left(\prod\limits_{i}^{}{\cal D}h_{\mu\nu}^i\right)
\exp\Biggl\{-\int d^4x\Biggl[\frac{1}{24\eta^2}
\left(H_{\mu\nu\lambda}^i\right)^2+\frac{Ng_m^2}{8}\left(h_{\mu\nu}^i\right)^2
-i\pi h_{\mu\nu}^i\Sigma_{\mu\nu}^{i{\,}(\alpha)}\Biggr]\Biggr\}=$$

$$
=\exp\left\{-2\pi^2\int d^4xd^4yD_m(x-y)\left[\eta^2\Sigma_{\mu\nu}^{i{\,}(\alpha)}(x)\Sigma_{\mu\nu}^{i{\,}(\alpha)}(y)
+\frac{2}{g_m^2}\frac{N-1}{N}j_\mu(x)j_\mu(y)\right]\right\},$$
where $D_m(x)=mK_1(m|x|)/(4\pi^2|x|)$ is the massive propagator with $K_1$ standing
for the modified Bessel function. Simplifying the integral
$\int d^4xd^4y\Sigma_{\mu\nu}^{i{\,}(\alpha)}(x)D_m(x-y)\Sigma_{\mu\nu}^{i{\,}(\alpha)}(y)$ (see ref.~\cite{theta}
for the analogous transformations in the $SU(3)$-case) we eventually arrive at the following final expression for the partition function:

$$
{\cal Z}_\alpha=
\exp\left\{-\frac{N-1}{4N}\left[g^2+\left(\frac{\Theta g_m}{\pi}\right)^2\right]
\int d^4x d^4y j_\mu(x)D_m(x-y)j_\mu(y)\right\}
\int\left(\prod\limits_{i}^{}{\cal D}x^{(i)}(\xi)\right)
\times$$

$$
\times\delta\left(\sum\limits_{i}^{}\Sigma_{\mu\nu}^i\right)
\exp\Biggl[
-2(\pi\eta)^2\int d^4x d^4y\hat\Sigma_{\mu\nu}^i(x)
D_m(x-y)\hat\Sigma_{\mu\nu}^i(y)-2i\Theta s_i^{(\alpha)}\hat L\left(\Sigma_i,C\right)+
$$

\begin{equation}
\label{main}
+2i\Theta\int d^4xd^4y\left(\frac{N-1}{N}\tilde\Sigma_{\mu\nu}(x)-s_i^{(\alpha)}\tilde\Sigma_{\mu\nu}^i(x)\right)j_\mu(y)
\partial_\nu^xD_m(x-y)\Biggr],
\end{equation}
where $\hat\Sigma_{\mu\nu}^i\equiv\Sigma_{\mu\nu}^i-Ns_i^{(\alpha)}\Sigma_{\mu\nu}$.
This formula is the main result of the present letter.
Note that
for every color $\alpha$, it is straightforward to
integrate out one of
the world sheets $\Sigma_i$'s by resolving the constraint
imposed by the $\delta$-function.

The first exponent
on the r.h.s. of eq.~(\ref{main}) represents the short-ranged interaction
of quarks via dual vector bosons. Noting that for any $\alpha$,
${\bf m}_\alpha^2=(N-1)/(2N)$, we immediately read from this term the total charge of the
quark, $\sqrt{g^2+(\Theta g_m/\pi)^2}$. The magnetic part of this charge coincides
with the one following from eq.~(\ref{ch}).
Further, the first term in the
second exponent on the r.h.s. of eq.~(\ref{main}) is the
short-ranged (self-)interaction of
closed world sheets $\Sigma_i$'s and an open one $\Sigma$.
In particular, by virtue of the general formulae obtained in ref.~\cite{mpla}, one can get from
the $\Sigma\times\Sigma$-interaction
the following values of the string tension and of the
inverse coupling constant of the rigidity term, corresponding to the confining-string world sheet $\Sigma$:

$$
\sigma=2\pi(N-1)\eta^2\ln\frac{m_H}{m},~~\alpha^{-1}=-\frac{\pi(N-1)}{4g_m^2N}={\cal O}\left(\frac{1}{N}\right).$$
Here, in the derivation of $\sigma$, we have in the standard way~\cite{ano} set for a characteristic small
dimensionless quantity in the model under study the ratio $m/m_H$ and adapted the logarithmic accuracy, i.e.,
assumed that not only $\frac{m_H}{m}\gg1$, but also $\ln\frac{m_H}{m}\gg1$. While the $1/N$ behavior
of $\alpha^{-1}$ is fixed by the requirement that $g_m^2\sim N$,
the $N$-dependence of $\sigma$ is subject to such a dependence of $\eta$. In QCD, to the leading order in the
parameter of the strong-coupling expansion, $\beta=2N/g^2$, the string tension for the rectangular
loop is known to be $N$-independent: $\sigma_{\rm QCD}=\frac{1}{a^2}\ln\frac{2N^2}{\beta}=\frac{1}{a^2}\ln\bar\lambda$,
where $a$ is the lattice spacing~\footnote{This fact stems also from the natural conjecture that the linear term in the
quark-antiquark potential should have the same $N$-dependence as the Coulomb term, that is
$V_{\rm Coul}(R)=-\frac{g_{\rm QCD}^2}{4\pi R}\frac{N^2-1}{2N}={\cal O}\left(N^0\right)$.}. Thus, if we adjust the $N$-dependence of $\eta$
as $\eta\sim\left[(N-1)\ln\frac{\sqrt{\lambda}}{N}\right]^{-1/2}$, where the $N$-dependence of $\lambda$ was discussed
in the paragraph following after eq.~(\ref{ch}), then the resulting string tension will be as $N$-independent,
as it is in QCD.

Next, the last term on the r.h.s. of eq.~(\ref{main}) describes
the short-range interactions
of dyons with both closed and open strings (obviously, the latter confine these very dyons themselves).
Finally, the term $-2i\Theta s_i^{(\alpha)}\hat L\left(\Sigma_i,C\right)$ in eq.~(\ref{main})
describes the long-range
interaction of dyons with closed world sheets, that is the 4D analogue of the Aharonov-Bohm effect~\cite{four}.
Since nonvanishing values of $s_i^{(\alpha)}$'s were found to be $\pm N^{-1}$,
at $\Theta\ne N\pi\times{\,}{\rm integer}$,
dyons (due to their magnetic charge) do interact by means of this term with the closed dual strings.
On the contrary, these critical values of $\Theta$
correspond to such a relation
between the magnetic charge of a dyon and the electric flux inside the
string when the scattering of dyons over strings is absent.
Note finally once more that these critical values of $\Theta$ generalize the $SU(2)$- and $SU(3)$-ones
obtained in refs.~\cite{emil} and~\cite{theta}, respectively.

\section{Acknowledgments}

The author is grateful for useful discussions to Prof. Adriano~Di~Giacomo and Drs. Cristina~Diamantini
and Luigi~Del~Debbio.
Besides that, he is grateful to
Prof. Adriano~Di~Giacomo and to the whole staff of the Physics Department of the
University of Pisa for cordial hospitality.
The work has been supported by INFN and partially by
the INTAS grant Open Call 2000, Project No. 110.


\newpage


\begin{thebibliography}{99}

\bibitem{th}
G. 't Hooft, Nucl. Phys. {\bf B 190} (1981) 455.

\bibitem{digiacomo}
J.M. Carmona, M. D'Elia, L. Del Debbio, A. Di Giacomo, B. Lucini, and G. Paffuti, {\it Color confinement
and dual superconductivity in full QCD}, hep-lat/0205025; A. Di Giacomo, {\it Color confinement and dual
superconductivity: an update}, hep-lat/0204032.

\bibitem{suN}
L. Del Debbio, A. Di Giacomo, B. Lucini, and G. Paffuti, {\it Abelian projection in SU(N) gauge theories},
hep-lat/0203023;
A. Di Giacomo, {\it Independence on the Abelian projection of monopole condensation in QCD}, hep-lat/0206018.

\bibitem{suNN}
M.C. Diamantini and C.A. Trugenberger, Phys. Rev. Lett. {\bf 88} (2002) 251601; JHEP {\bf 04} (2002) 032;
U. Ellwanger and N. Wschebor, JHEP {\bf 10} (2001) 023; D. Antonov, Mod. Phys. Lett. {\bf A 17} (2002) 279.

\bibitem{dw}
S.R. Das and S.R. Wadia, Phys. Rev. {\bf D 53} (1996) 5856.

\bibitem{tHM}
S. Mandelstam, Phys. Lett. {\bf B 53} (1975) 476; Phys. Rep.
{\bf 23} (1976) 245; G. 't Hooft, in: {\it High Energy Physics},
Ed. A. Zichichi (Editrice Compositori, Bologna, 1976).

\bibitem{pb}
A.M. Polyakov, {\it Gauge fields and strings} (Harwood Academic Publishers, Chur, 1987).

\bibitem{maedan}
S. Maedan and T. Suzuki, Prog. Theor. Phys. {\bf 81} (1989) 229.

\bibitem{su3}
D. Antonov and D. Ebert, Phys. Lett. {\bf B 444} (1998) 208;
D.A. Komarov and M.N. Chernodub, JETP Lett. {\bf 68} (1998) 117.

\bibitem{theta}
D. Antonov, Phys. Lett. {\bf B 475} (2000) 81.

\bibitem{moresu3}
D. Antonov, Mod. Phys. Lett. {\bf A 14} (1999) 1829; JHEP {\bf 07} (2000) 055;
Nucl. Phys. (Proc. Suppl.) {\bf B 96} (2000) 491.

\bibitem{ano}
A.A. Abrikosov, Sov. Phys. JETP {\bf 5} (1957) 1174;
H.B. Nielsen and P. Olesen, Nucl. Phys. {\bf B 61} (1973) 45;
for a review see e.g.: E.M. Lifshitz and L.P. Pitaevski,
{\it Statistical Physics, Vol. 2} (Pergamon, New York, 1987).

\bibitem{emil}
E.T. Akhmedov, JETP Lett. {\bf 64} (1996) 82.

\bibitem{witten}
E. Witten, Phys. Lett. {\bf B 86} (1979) 283.

\bibitem{polikarp}
E.T. Akhmedov, M.N. Chernodub, M.I. Polikarpov, and M.A. Zubkov,
Phys. Rev. {\bf D 53} (1996) 2087.

\bibitem{group}
R. Gilmore, {\it Lie groups, Lie algebras, and some of their applications}
(J. Wiley \& Sons, New York, 1974).

\bibitem{mpla}
D.V. Antonov, D. Ebert, and Yu.A. Simonov, Mod. Phys. Lett. {\bf A 11} (1996) 1905.

\bibitem{four}
M.G. Alford, J. March-Russel, and F. Wilczek, Nucl. Phys.
{\bf B 337} (1990) 695; J. Preskill and L.M. Krauss, Nucl. Phys.
{\bf B 341} (1990) 50.







\end{thebibliography}
\end{document}